\documentclass[twocolumn,prb,aps]{revtex4}
\usepackage{amsmath}
\usepackage{amssymb}
\usepackage{graphicx}
\usepackage{dcolumn}
\usepackage{graphicx}
\usepackage{dcolumn}
\usepackage{bm}

\begin{document}

%\preprint{Phys.Rev.B}

\title{Thermopower of a Two-Dimensional Semimetal in a HgTe Quantum Well}
\author{E. B. Olshanetsky,$^1$ Z. D. Kvon,$^{1,2}$ M. V.~Entin,$^1$, L. I.~Magarill,$^1$ A. Levin$^3$ G. M. Gusev,$^3$ N. N. Mikhailov$^1$}

\affiliation{$^1$Institute of Semiconductor Physics, Novosibirsk
630090, Russia}

\affiliation{$^2$Novosibirsk State University, Novosibirsk 630090,
Russia}

\affiliation{$^3$Instituto de F\'{\i}sica da Universidade de S\~ao
Paulo, 135960-170, S\~ao Paulo, SP, Brazil}

\date{\today}
\begin{abstract}
The thermopower in a two-dimensional semimetal existing in HgTe
quantum wells 18–21 nm thick has been studied experimentally and
theoretically for the first time. It has been found theoretically
and experimentally that the thermopower has two
components—diffusion and phonon-drag—and that the second component
is several times larger than the first. It has been concluded that
the electron–hole scattering plays an important role in both
mechanisms of the thermopower.


\end{abstract}

\maketitle

A two-dimensional semimetal appearing in HgTe quantum wells 18–21
nm thick [1, 2] currently attracts permanent interest because it
is a two-component electron–hole system with a number of unusual
properties [3–12] caused by the coexistence of electrons and
holes. One of these properties is electron–hole scattering through
the Landau mechanism, which is responsible for a strong
temperature dependence of the resistance of a two-dimensional
semimetal, in contrast to a single-component system. It is
obviously important to comprehensively study kinetic effects in
this system.

In this work, we report the first experimental study of the
thermopower of a two-dimensional semimetal. The comparison of the
experiment with the theoretically predicted diffusion contribution
to the thermopower in the presence of electron–hole scattering
shows that this contribution underestimates the thermopower.
Consequently, it is necessary to take into account another
possible contribution to the thermopower from phonon drag of
electrons and holes.

We studied $4\times3$-mm rectangular samples whose middle parts
contain Hall bars with $L \times W = 100 \times50 - \mu$m and $250
\times 50 - \mu m$ segments based on wide HgTe quantum wells 20 nm
thick with the (013) orientation. The thermopower was measured as
follows. A heater in the form of a thin metallic strip with the
resistance $\approx 100 \Omega $ was placed on one side of a
sample against one of the electric contacts (see the inset of Fig.
1b). The opposite end of the sample through a deposited indium
layer was in thermal contact with a $5 - mm ^ { 3 }$ copper
thermal anchor, which was in turn in contact with a massive copper
holder of the sample. To create a temperature gradient along the
sample, an alternating current with a frequency of 0.4–1 Hz and a
magnitude of no more than 60 mA was passed through the metallic
strip (heater). The heater operated in a linear regime in the
indicated current range. The temperature gradient appearing along
the sample was controlled using two calibrated thermistors placed
on the sides of the heater and thermal anchor. In particular, the
temperature difference thus determined between contacts spaced
from each other by a distance of 100$\mu m$ was $\Delta T \approx
0.023$ K at T=4.2 K and $V _ { \mathrm { Heat } } =$ 6 V. The
thermal conductivity of liquid helium in the working temperature
range ($\approx 2.2 - 4.2$ K) was negligibly low as compared to
the phonon thermal conductivity of the substrate. Under these
conditions, the thermal conductivity of the substrate determines
the temperature gradient along the sample. The thermopower signal
was measured at the double frequency with the use of all
potentiometric contacts. We studied about ten samples.

\begin{figure}[h]
%h=here, t=top, b=bottom, p=separate figure page
\begin{center}\leavevmode
\includegraphics[width=0.9\linewidth]{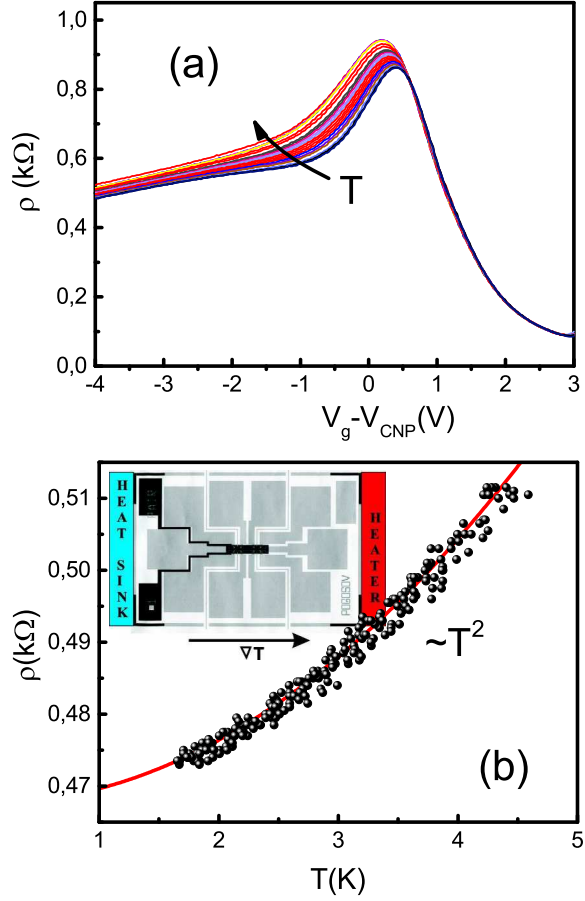}
\caption{Fig. 1 (Color online)(a) Resistance of the structure
versus the gate voltage at various temperatures in the range of $T
= 2.1 - 6 \mathrm { K }$. (b) Temperature dependence of the
resistance of the sample at $V _ { \mathrm { g } } = - 5 \mathrm {
V }$. Here and below, $V _ { \mathrm { CNP } } = - 1.1 \mathrm { V
}$ is the charge neutrality point. The inset shows the structure
under study and the direction of the temperature gradient created
by the heater (on the right) and heat sink (on the left).}
\end{center}
\end{figure}

We begin the description of the experiment with the analysis of
the transport response of the studied samples. Figure 1à shows
typical dependences of the resistance on the gate voltage at
different temperatures. It is seen that these dependences
correspond to the behavior expected for 20-nm HgTe quantum wells
where a (two-dimensional metal–two-dimensional semimetal)
transition occurs at the variation of the gate voltage [2, 13].
This transition is accompanied by a sharp change in the
temperature dependence of the resistance: this dependence is very
weak before the transition and represents a typical temperature
dependence of a two-dimensional metal at $k_{F}l >> l$ ($k_F$ and
$l$ - are the wave vector and mean free path of the electron) and
low temperatures when the phonon Seebeck coefficient is almost
absent and the temperature dependence is determined by
weak-localization effects, whereas a noticeable increase in the
resistance with the temperature is observed after the transition
to the semimetal state, which is due to electron–hole scattering
and is thereby proportional to the temperature squared (Fig. 1b).

\begin{figure}[h]
%h=here, t=top, b=bottom, p=separate figure page
\begin{center}\leavevmode
\includegraphics[width=0.9\linewidth]{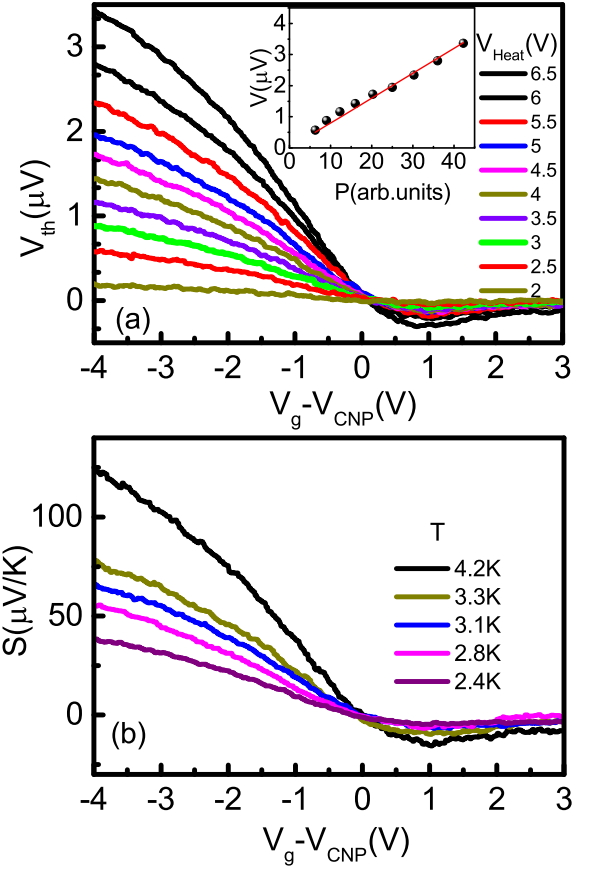}
\caption{Fig. 2. (Color online) (a) Thermopower versus the gate
voltage for various heater powers at the temperature $T = 4.2
\mathrm { K }$. The inset shows the thermopower versus the heater
power at $V _ { \mathrm { g } } = - 5 \mathrm { V }$. (b) Seebeck
coefficient versus the gate voltage at various temperatures.}
\end{center}
\end{figure}

Figure 2a shows the gate voltage dependences of the
temperature-gradient-induced potential difference $V_{th}$ between
the potentiometric contacts of the bar with the length $L=100 \mu
m$. The inset of this figure shows the dependence of the
thermopower signal on the power applied to the heater. It is
clearly seen that the measured signal is proportional to this
power, which indirectly indicates that the measured signal is
indeed due to the thermopower rather than possible pickups. We now
qualitatively analyze the behavior of the thermopower shown in
Fig. 2a. We begin with the dependence on the gate voltage. At gate
voltages corresponding to the electron metal, the thermopower is
relatively low and decreases with an increase in the density
according to the Mott formula for the thermopower of metals. The
thermopower changes sign near the transition point and begins to
increase almost linearly with the development of the semimetal
state (with an increase in the density of holes). Figure 2b shows
the gate voltage dependences of the Seebeck coefficient $S = V _ {
\mathrm { th } } / \Delta T ( \Delta T$ is the temperature
difference between potential contacts on which the signal is
measured) at different temperatures. It is seen that the Seebeck
coefficient increases with the temperature of the sample.

To more accurately describe the experimental results obtained in
this work, we developed a theory of the diffusion component of the
thermopower for a two-dimensional electron–hole system consisting
of two types of degenerate particles, electrons and holes. In the
presence of the temperature gradient, chemical potential, and
electron–hole friction, the average velocities in the subsystems
satisfy the equation

\begin{eqnarray} n _ { \mathrm { v } } e _ { \mathrm { v } }
\mathbf { E } & - g _ { \mathrm { v } } m _ { \mathrm { v } }
\frac { \pi T } { 3 \hbar ^ { 2 } } \nabla T - \frac { m _ {
\mathrm { v } } n _ { \mathrm { v } } } { \tau _ { \mathrm { v } }
} \mathbf { V } _ { \mathrm { v } } \nonumber \\ & = \eta n _ {
\mathrm { v } } n _ { \overline { \mathrm { v } } } \left( \mathbf
{ V } _ { \mathrm { v } } - \mathbf { V } _ { \overline { \mathrm
{ v } } } \right) \end{eqnarray}

Here, the subscript $\mathrm { v } = ( e , h )$ specifies
quantities referring to electrons (e) and holes (h); $n _ { e }$
($ n _ { h }$) is the density of electrons (holes); $g _ { \mathrm
{ v } }$ is the number of valleys ($g _ { e } = 1 , g _ { h } =
2$); $\mathbf { V } _ {\mathrm { v }} , m _ { \mathrm { v }}$, and
$e _ { \mathrm { v } }$ are the average velocity, effective mass,
and charge of particles of type $\mathrm { v }$ ($e _ { e , h } =
\mp e$ , e is the charge of the hole), respectively; $\tau _ {
\mathrm { v } }$ is the relaxation transport time on impurities;
$\eta$ is the friction coefficient; and T is the temperature in
energy units. The friction coefficient $\eta = \Theta T ^ { 2 }$
is determined by electron–hole scattering through the Landau
mechanism. Equation (1) is a generalization of equations presented
in [10, 11] to the case of the existence of a temperature
gradient.

From the condition that the total current density $\mathbf { j } =
e \left( - n _ { e } \mathbf { V } _ { e } + n _ { h } \mathbf { V
} _ { h } \right)$ vanishes, the Seebeck coefficient is obtained
in the form

\begin{eqnarray}
S = - \frac { \pi } { 3 e \hbar ^ { 2 } } T\times \nonumber
\\ \frac { m _ { e } m _ { h } \left( g _ { e } \tau _ { e } - g _
{ h } \tau _ { h } \right) + \eta \tau _ { e } \tau _ { h } \left(
g _ { e } m _ { e } + g _ { h } m _ { h } \right) \left( n _ { e }
- n _ { h } \right) } { m _ { h } n _ { e } \tau _ { e } + m _ { e
} n _ { h } \tau _ { h } + \left( n _ { e } - n _ { h } \right) ^
{ 2 } \eta \tau _ { e } \tau _ { h } }
\end{eqnarray}

It is noteworthy that the contribution to the current from any
type of charge carriers in Eq. (2) is nonzero even at zero carrier
density; i.e., this formula does not have any monopolar limit:

\begin{eqnarray}
S _ { e , h } ^ { ( 0 ) } = \mp \frac { \pi } { 3 e \hbar ^ { 2 }
} T \frac { m _ { e , h } g _ { e , h } } { n _ { e , h } }
\end{eqnarray}

In contrast to Eq. (2), Eq. (3) does not include terms
corresponding to the second type of carriers, in particular, its
relaxation time and friction. The reason for this between formulas
(2) and (3) are different because they are obtained under the
assumption that Fermi gases are degenerate. Indeed, the transition
to the monopolar case at low temperatures occurs in a relatively
narrow range of the chemical potential $\Delta \zeta \sim T$. The
friction between different types of carriers distorts the linear
temperature dependence of S. In the low-temperature limit, $\eta
\propto T ^ { 2 }$ , which leads to third-order temperature
corrections to the linear dependence.

The friction can become a prevailing mechanism of scattering
($\eta \rightarrow \infty$ ) at higher temperatures. In this case,
Eq. (2) becomes independent of all relaxation constants:

\begin{eqnarray}
S = - \frac { \pi } { 3 e \hbar ^ { 2 } } T \frac { m _ { e } g _
{ e } + m _ { h } g _ { h } } { n _ { e } - n _ { h } }
\end{eqnarray}

This formula is valid far from the charge neutrality point (CNP).
The Seebeck coefficient S changes sign near this point (more
precisely, at the point where the numerator of Eq. (2) changes
sign). The Seebeck coefficient in the region of applicability of
Eq. (4) also has a linear temperature dependence similar to that
at low temperatures but with a larger slope. As a result, the
dependence can be close to a quadratic law in the intermediate
temperature range.

Figure 3a shows the Seebeck coefficients $S \left( V _ { g }
\right)$ calculated by Eqs. (2) and (3) in comparison with
experimental data. All parameters necessary for the calculation by
Eqs. (2) and (3) (mobilities and densities of electrons and holes
and the friction coefficient) and their dependence on the gate
voltage were obtained previously from transport measurements [11]
and from the cyclotron resonance (effective masses of electrons
and holes) [14]. The temperature gradient necessary for the
determination of the Seebeck coefficient was measured
experimentally according to the method described at the beginning
of this paper. Thus, the comparison of the theory and experiment
in Fig. 3a is free of fitting parameters.

The qualitative behavior of the Seebeck coefficient to the right
of the charge neutrality point, where the electron metal exists,
corresponds to the Mott theory for metals, which predicts a
decrease in the Seebeck coefficient with an increase in the
carrier density. For comparison with experimental data in this
range of gate voltages, we used the monopolar formula (3) for
electrons, which is the Mott formula under the assumption that
$\tau ( \varepsilon ) = \mathrm { const }$ ($\tau$ is the pulse
relaxation time and $\varepsilon$ is the energy). As is seen, the
Seebeck coefficients calculated by Eq. (3) are in satisfactory
agreement with experimental points (Fig. 3a).

\begin{figure}[h]
%h=here, t=top, b=bottom, p=separate figure page
\begin{center}\leavevmode
\includegraphics[width=0.9\linewidth]{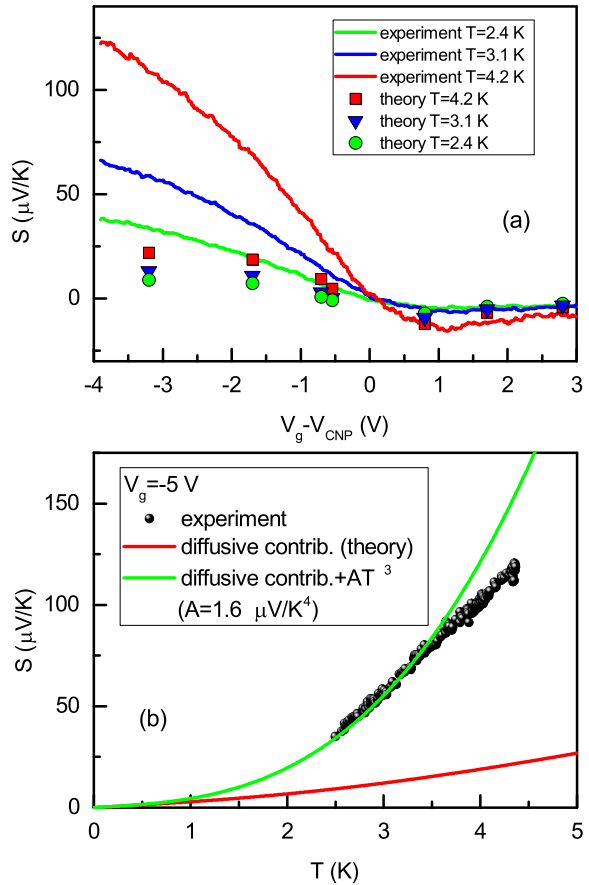}
\caption{Fig. 3. (Color online) (a) Seebeck coefficient versus the
gate voltage at the temperatures T=2.4, 3.1, and 4.2 K according
to (lines) experiments and (points) calculations by Eqs. (2) and
(3) to the left and right of the charge neutrality point,
respectively, with the parameters obtained from transport
measurements. (b) Temperature dependence of the Seebeck
coefficient in the two-dimensional semimetal: (points) experiment
for $V_g=-5$V; (lower red line) diffusion contribution that
corresponds to the indicated gate voltage and is calculated by Eq.
(2); and (upper green line passing through experimental points)
sum the diffusion contribution shown by the lower red line and the
function $S = A T ^ { 3 }$ ($A = 1.6 \mu \mathrm { V } / \mathrm {
K } ^ { 4 }$), which represents the assumed phonon drag
contribution.}
\end{center}
\end{figure}

On the contrary, to the left of the charge neutrality point in
Fig. 3a, i.e., in the region of gate voltages corresponding to the
formation of the two-dimensional semimetal, agreement between the
experiment and theory (Eq. (2)) is much worse. In this range of
gate voltages, the theory gives Seebeck coefficients about
one-fourth of experimental values (see Fig. 3a). This discrepancy
apparently appears because Eq. (2) describes only the diffusion
contribution to the thermopower of the semimetal. However, the
measured thermopower can include not only the diffusion
contribution but also the phonon drag contribution [15], which is
disregarded in our theory. As is known, the phonon drag is
proportional to the mass squared of charge carriers. The masses of
electrons and holes in the 20-nm HgTe quantum well are $m _ { e }
= 0.025 m _ { 0 }$ and $m _ { h } = 0.15 m _ { 0 }$, respectively.
For this reason, the phonon drag contribution on the left of the
charge neutrality point (i.e., in the region where holes dominate)
is significant, whereas this contribution on the right of the
charge neutrality point, where the two-dimensional metal exists,
is negligible.

Figure 3b shows the experimental points for the temperature
dependence of the Seebeck coefficient at $V _ { g } = - 5 V$. It
is seen that the experimental data are significantly higher than
the corresponding diffusion contribution calculated by Eq. (2)
shown by the lower red line in Fig. 3b. It can be assumed that the
difference between the shown experimental and calculated
dependences corresponds to the contribution to the Seebeck
coefficient from phonon drag in the twodimensional semimetal under
the condition of dominance of holes. As an example of phonon drag
in an ordinary two-dimensional metal, we consider the contribution
$\sim T ^ { 3 }$ [15]. Through the experimental points in Fig. 3b,
we plot the upper green line representing the sum of the diffusion
contribution and the function $S = A T ^ { 3 }$ ($A = 1.6 \mu
\mathrm { V } / \mathrm { K } ^ { 4 }$). It is seen that this line
reproduces well the experimental data in the range of 2.5–3.5 K,
but a discrepancy is observed at higher temperatures, where the
experimental points are below the calculated curve. This
discrepancy can be attributed to the scattering of phonon-dragged
holes by electrons, which is enhanced with increasing temperature,
reducing the measured Seebeck coefficient. However, for a more
definite conclusion, it is necessary to develop a theory of phonon
drag in the twodimensional semimetal in the presence of electron–
hole scattering.

To summarize, experimental information on the behavior of the
thermopower in a two-dimensional semimetal has been obtained for
the first time. A theory of the diffusion component of the
thermopower in the two-dimensional semimetal in the presence of
electron–hole scattering has been developed. This theory
underestimates the experimentally observed Seebeck coefficients.
This discrepancy indicates the necessity of the inclusion of the
electron–phonon drag in the two-dimensional semimetal in the
presence of the electron–hole scattering.

This work was supported by the Russian Science Foundation (project
no. 16-12-10041). The work of L.I.M. and M.V.E. was supported by
the Russian Foundation for Basic Research (project no. 17-02-
00837).

\end{document}